\documentclass[preprint,superscriptaddress,preprintnumbers,amsmath,amssymb,pra]{revtex4}
\usepackage{graphicx}
\usepackage{amsmath}

\begin{document}
\pdfoutput=1
\thispagestyle{empty}

\title{Precision measurements of the gradient of the Casimir force between
ultra clean metallic surfaces at larger separations}

\author{Mingyue Liu}
\affiliation{Department of Physics and Astronomy, University of California, Riverside, California 92521, USA}

\author{Jun Xu}
\affiliation{Department of Physics and Astronomy, University of California, Riverside, California 92521, USA}

\author{
G.~L.~Klimchitskaya}
\affiliation{Central Astronomical Observatory at Pulkovo of the
Russian Academy of Sciences, Saint Petersburg,
196140, Russia}
\affiliation{Institute of Physics, Nanotechnology and
Telecommunications, Peter the Great Saint Petersburg
Polytechnic University, Saint Petersburg, 195251, Russia}

\author{
V.~M.~Mostepanenko}
\affiliation{Central Astronomical Observatory at Pulkovo of the
Russian Academy of Sciences, Saint Petersburg,
196140, Russia}
\affiliation{Institute of Physics, Nanotechnology and
Telecommunications, Peter the Great Saint Petersburg
Polytechnic University, Saint Petersburg, 195251, Russia}
\affiliation{Kazan Federal University, Kazan, 420008, Russia}

\author{
 U.~Mohideen\footnote{Umar.Mohideen@ucr.edu}}
\affiliation{Department of Physics and Astronomy, University of California, Riverside, California 92521, USA}

\begin{abstract}
We report precision measurements of the Casimir interaction at larger separation
distances between the Au-coated surfaces of a sphere and a plate in
ultrahigh vacuum using a much softer cantilever of the dynamic atomic force
microscope-based setup and two-step cleaning procedure of the vacuum chamber and
test body surfaces by means of UV light and Ar-ion bombardment.
Compared to the previously performed experiment, two more measurement sets
for the gradient of the Casimir force are provided which confirmed and
slightly improved the results. Next, additional
measurements have been performed with a factor of two larger
oscillation amplitude of the cantilever. This allowed obtaining meaningful
results at much larger separation distances. The comparison of the
measurement data with theoretical predictions of the Lifshitz theory
using the dissipative Drude model to describe the response of Au to the
low-frequency electromagnetic field fluctuations shows that this theoretical approach
is experimentally excluded over the distances from 250 to 1100~nm
(i.e., a major step forward has been made as compared to the previous
work where it was excluded up to only 820~nm). The theoretical approach
using the dissipationless plasma model at low frequencies is shown to be
consistent with the data over the entire measurement range from 250 to
1300~nm. The possibilities to explain these puzzling results are discussed.
\end{abstract}

\maketitle

\section{Introduction}

Extensive studies of the Casimir force in the last two decades
lead us to conclude that this fluctuation-induced quantum phenomenon is
of considerable importance for both fundamental physics and its
technological applications (see the monograph \cite{1} and
reviews \cite{2,3,4}). For almost half a century it was
generally believed that the Lifshitz theory \cite{5} provides quite a
satisfactory description of the van der Waals and Casimir forces acting
between the closely spaced surfaces made of various materials.
In so doing the single input parameter needed to make theoretical
predictions was the frequency-dependent dielectric permittivity
of the interacting bodies describing their response to the electromagnetic
field. Contrary to expectations, several precise measurements
performed in the last fifteen years resulted in contradictions
between experiment and theory which are sometimes called the Casimir
puzzle and Casimir conundrum to specify the problems arising for
metallic and dielectric or semiconductor materials,
respectively \cite{5a,5b}.

The Casimir puzzle consists in the fact that theoretical predictions
of the Lifshitz theory for metallic test bodies obtained with inclusion
of the relaxation properties of conduction electrons are excluded by the
measurement data of all precise experiments at short separations
\cite{6,7,8,9,10,11,12,13,14,15}.
The dielectric response of metals used in computations is found from
the optical data extrapolated down to zero frequency by means of the
Drude model, where the relaxation parameter $\gamma$ describes the
energy losses of conduction electrons.
It is puzzling also that if one puts $\gamma$ equal to zero (as if
there were no energy losses at low frequencies) the Lifshitz theory
comes to good agreement with the measurement data of the
same experiments \cite{6,7,8,9,10,11,12,13,14,15} (recall that all
of them have been performed at separations below 750~nm between the
interacting bodies). This means that the dissipationless plasma model, which
is in fact applicable only at high frequencies in the region of
infrared optics, works well for some reasons even at low frequencies
characteristic of the normal skin effect. The problem is aggravated
by the fact that for metals with perfect crystal lattices the Casimir
entropy calculated within the Lifshitz theory using the Drude model
violates the Nernst heat theorem although the same satisfies it if the plasma
model is used \cite{16,17,18,19,20,21}.

In a similar way, the term Casimir conundrum is applied in reference
to the fact that theoretical predictions of the Lifshitz theory for
dielectrics and dielectric-type semiconductors obtained with the
inclusion of the dc conductivity  are excluded by the
measurement data of several experiments \cite{23,24,25,26,27}.
If the conductivity of dielectric materials at room temperature is
disregarded, the Lifshitz theory comes to agreement with the same
data \cite{22,23,24,25,26,27}. By analogy with the case of metals,
it has been proven that the Casimir entropy calculated in the framework
of the Lifshitz theory violates the Nernst heat theorem if the
conductivity of dielectrics is taken into account which is
otherwise satisfied  \cite{28,29,30}. Considering that all dielectric
materials possess a rather small but nonzero electric conductivity
at any nonzero temperature, the situation should be considered as
paradoxical.

Resolution of the problems arising in the Lifshitz theory for both
metallic and dielectric materials is of major importance for
applications of the Casimir force in nanotechnology.
With the decrease in separations between the moving parts of
microelectromechanical devices below a micrometer, the Casimir
force becomes dominant. Because of this, it has long been proposed
\cite{31} that the next generation of micro- and nanodevices will
exploit the Casimir force for their functionality.
In the early twenty first century, extensive studies of the role of
Casimir force in microdevices have been conducted and a lot of
devices driven by the Casimir force, such as oscillators, switches,
microchips etc., have been proposed
\cite{32,33,34,35,36,37,38,39,40,41,42,43,44,45,46}.
The description of their functionality on the basis of the Lifshitz
theory essentially depends on whether the dissipative Drude or the
dissipationless plasma model is used
in extrapolation of the optical data to low frequencies.
This places strong emphasis on the Casimir puzzle and Casimir
conundrum as they impact
fundamental physics as well as technology.

All precision measurements of the Casimir interaction mentioned above lead
to meaningful results at relatively short separations below 750~nm
between the test bodies (with the single exception of measuring the
Casimir-Polder force \cite{22} relevant to the Casimir conundrum).
After the experiment \cite{14} on measuring the difference Casimir
force was performed, where the theoretical predictions of the Lifshitz
theory using the Drude and plasma models differ by up to a factor of
1000, an exclusion of the former within this separation range was
conclusively established. In so doing the role of different background
effects, such as the surface roughness \cite{47,48}, variations in
the optical data \cite{49}, patch potentials \cite{50,51}  etc., as well
as the validity of calculation procedure (including the
role of deviations from
the proximity force approximation \cite{53a,52,53,54,55}), were investigated
in detail.

Special attention was paid to the background forces due to patch
potentials which become much larger than the Casimir force at separations
of a few micrometers. Thus, in Ref.~\cite{56} an attempt was undertaken to
extract the Casimir force from up to an order of magnitude larger forces
between a centimeter-size spherical lens and a plate, presumably caused by
the patch potentials, by means of some fitting procedure. The obtained
Casimir force was found to be in better agreement with the Drude model
approach. It was shown \cite{57}, however, that depending on uncontrolled
imperfections on the lens surface the obtained results may agree with equal
ease either with the Drude or the plasma model approaches.

Taking into account the significance of the above problems in
Casimir physics, which remain unsolved for almost twenty years, in Ref.~\cite{58}
an upgraded atomic force microscope (AFM)-based technique was developed and an
advanced surface cleaning procedure was used in order to eliminate the role of
patch potentials and make progress towards a precision measurement of the Casimir
interaction at larger separations. For this purpose both interior surfaces of
the vacuum chamber and Au-coated test bodies (the sphere and the plate) were
successively cleaned by means of UV light and Ar ions. Another improvement
was the use of a much softer cantilever. These allowed the separation-independent
and low residual potential difference as well as a sixfold decrease of the
systematic error in measuring the gradient of the Casimir force.
The measurement data have been compared with theoretical predictions of the
Lifshitz theory obtained using the extrapolations of the optical data of Au
to zero frequency by means of the Drude and plasma models. As a result,
the Drude model approach was excluded and the plasma model approach confirmed
by the data up to the sphere-plate separation distance of 820~nm.

In this paper, we continue the investigation of the Casimir force between an
Au-coated sphere and a plate at larger separation distances using an upgraded
technique and the cleaning procedure of Ref.~\cite{58}. The results reported
in the rapid communication \cite{58} were based on a single measurement set
(the gradient of the Casimir force was measured for 21 times at each separation
over the range from 250 to 950~nm with a step of 1~nm). Here, we discuss the
data of two additional measurement sets and present the mean results from all
the three sets. The comparison of these results with theoretical predictions
of the Lifshitz theory made using two different statistical procedures
leads to the exclusion of the Drude model approach and confirmation of
the plasma model approach up to the separation of 850~nm.

Next we present additional measurements with increased oscillation amplitude
of the cantilever (20~nm instead of 10~nm in Ref.~\cite{58}) which decreased
by the factor of 1.375 the systematic error in measuring the frequency shift.
The obtained measurement data are again compared with theoretical predictions
of the Lifshitz theory using the two alternative statistical approaches.
As a result, the Drude model approach is excluded
up to much larger separation distance of 1100~nm.
The plasma model approach is found to be
consistent with the data up to the separation of $1.3~\mu$m.
The importance and possibilities to test the Lifshitz theory experimentally
at even larger separations are discussed.

The paper is organized as follows. In Sec.~II we briefly present the upgraded
AFM-based setup and some additional details of the cleaning procedure by means
of UV light and Ai-ion bombardment. Section~III is devoted to the measurement
results with relatively small oscillation amplitude of the cantilever and their
comparison with theory. In Sec.~IV the measurement results at larger separations
are presented and compared with theoretical predictions. In Sec.~V the reader
will find our conclusions and a discussion of future prospects.

\section{Upgraded setup with a two-step surface cleaning}

We have measured the gradient of the Casimir force between an Au-coated hollow
glass sphere and an Au-coated polished Si wafer by means of the AFM-based setup
working in the frequency-shift mode in ultrahigh vacuum. The main steps in making
these measurements are the following.

\begin{figure}[t]
\vspace*{-6cm}
\centerline{\hspace*{1.5cm}
\includegraphics[width=6.50in]{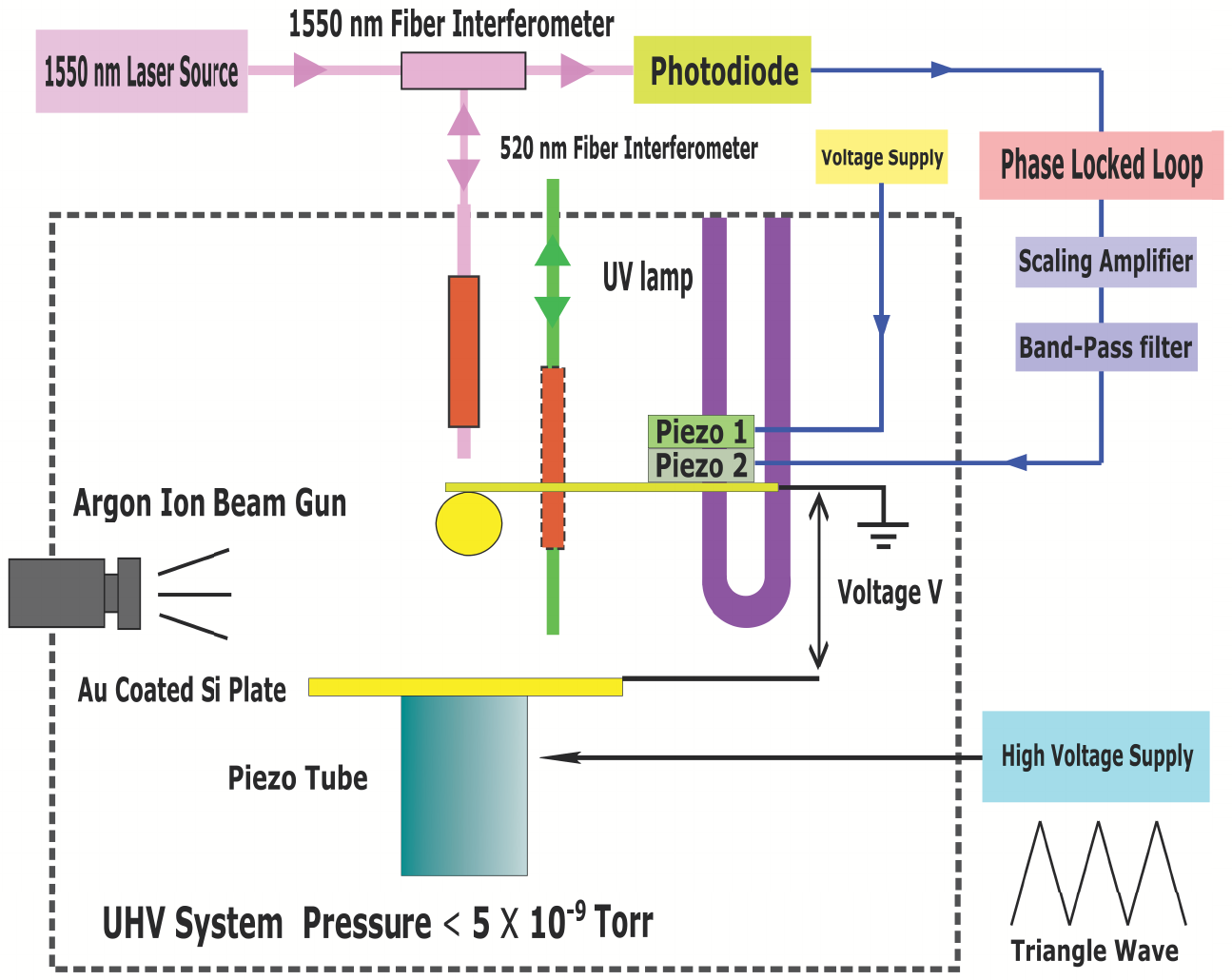}
}
\vspace*{-10.5cm}
\caption{\label{fg1}
Schematic of the upgraded experimental setup (see text for
further discussion) placed inside the vacuum chamber with
a pressure $<5\times 10^{-9}~\mbox{Torr}\approx 0.7\times 10^{-6}~$Pa.
}
\end{figure}

The force-sensitive element in our setup shown schematically in Fig.~\ref{fg1}
is a rectangular cantilever. As compared to previous experiments
\cite{10,11,12,13,15}, the precision of force measurements here was
improved by increasing the sensitivity of the cantilever through a decrease of its
spring constant. The cantilever spring constant is given by \cite{59}
\begin{equation}
k=\frac{wv^3Y}{4L^3},
\label{eq1}
\end{equation}
\noindent
where  $w,~v$, and $L$ are the width, thickness, and length of the cantilever beam,
respectively, and $Y$ is its Young's modulus. As is seen from Eq.~(\ref{eq1}),
the spring constant can be effectively decreased by reducing the
thickness of the beam.

\begin{figure}[b]
\vspace*{-6cm}
\centerline{\hspace*{1.5cm}
\includegraphics[width=6.50in]{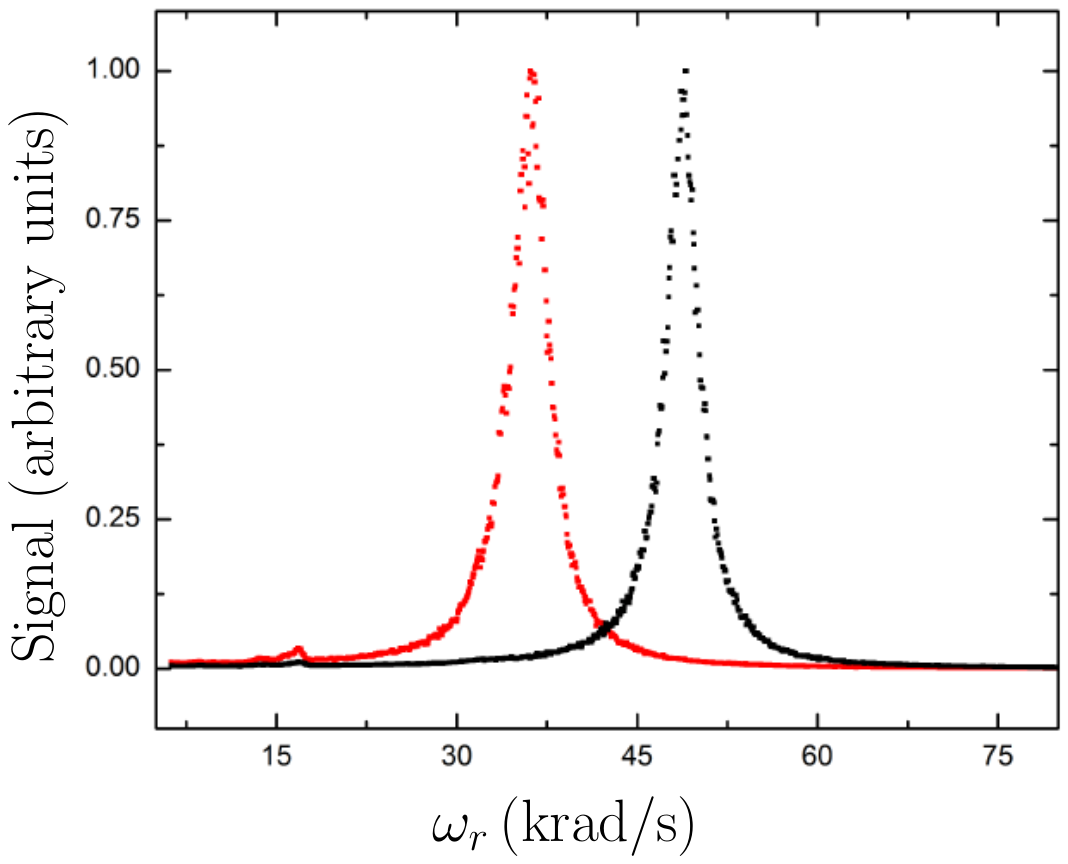}
}
\vspace*{-10.5cm}
\caption{\label{fg2}
The cantilever thermal noise oscillation spectrum  is shown as a function
of frequency before (right peak) and after (left peak) etching.
}
\end{figure}

This was achieved by means of the etching process. At first, the cantilever was
rinsed with buffered HF solution (BOE 6:1) for 1~min. followed by DI water to
remove the oxide layer. Then the cantilever was etched with 60\% KOH solution at
$T=50\,{}^{\circ}$C for 55~s. Mild agitation by hand was used to obtain a uniform
etching. Relatively high concentration of KOH solution and high temperature were
necessary to achieve sufficiently smooth surfaces after etching \cite{61}.
The spring constant was measured using the thermal oscillation spectrum of the
cantilever, as discussed in Ref.~\cite{60}, both before and after the etching
process and the values $k=0.013$ and 0.0063~N/m were obtained, respectively.
As a result, the resonant frequency of the cantilever $\omega_0$ was reduced
from its original value $4.877\times 10^4~$rad/s to $3.608\times 10^4~$rad/s
(see Fig.~\ref{fg2}). These measurements were made in ambient conditions at
room temperature.

The first test body of our setup is the hollow glass sphere of approximately
$43~\mu$m radius. It was made from liquid phase which leads to almost perfectly
spherical shape with less than 0.1\% relative difference along any two
perpendicular axes. The sphere was baked at $60\,{}^{\circ}$C for two hours
to remove volatile components. Then it was picked up using a bare
optical fiber and attached to a cantilever using a very small amount of
conducting silver epoxy (see Fig.~\ref{fg1}).
The process of attachment was performed under an optical microscope.

Following the attachment of the sphere, the Au coating was applied on the cantilever
and the sphere using an E-beam evaporator at a pressure
$5\times 10^{-6}~\mbox{Torr}\approx 0.7\times 10^{-3}~$Pa.
In contrast to thermal evaporators used in previous experiments, the E-beam
evaporator leads to smoother surfaces and lower roughness. The Au coating speed
was 2~\AA/s and the thickness of the Au layer was $118\pm 1~$nm.
After the measurements of the Casimir force were completed, the radius of the
Au-coated sphere was measured to be $R=43.466\pm0.042~\mu$m using a
calibrated scanning
electron microscope and software ImageJ to precisely determine the sphere
boundary.
Quantification of the
deviation from a perfect sphere was done by finding the difference between
the maximum and minimum diameters for any two perpendicular line scans of
the sphere diameter in ImageJ. The respective difference in the sphere
radii was taken into account in the total error of $R$ indicated above.
As a result, the error in determination of the sphere radius was decreased
 as compared to previously reported in the literature (see, e.g.,
Refs.~\cite{10,60a,60b}).
The rms roughness on the sphere surface $\delta_s=1.13~$nm
was also measured when the work was completed. After the Au coating, the spring
constant of a cantilever with sphere attached
increased to $k=0.007353$~N/m, and its resonant frequency
decreased to $\omega_0=0.9444\times 10^4~$rad/s in ultrahigh vacuum.

The Au-coated sphere-cantilever system was attached to two block piezoelectric
actuators. The cantilever is electrically grounded (see Fig.~\ref{fg1}).
The cantilever motion was monitored using a fiber optical interferometer with
a laser light wavelength of 1550~nm. For so doing, single mode 1550~nm fibers
were used. To improve the finesse of the Fabry-Perot cavity of the interferometer,
the reflectance of the top side of the cantilever was increased with a layer
of Au of 40~nm thickness.

The second test body of our setup is a polished Si wafer of
$1\times 1~\mbox{cm}^2$ area and of $500~\mu$m thickness used as a plate.
It can be considered infinitely large as compared to the sphere (in Fig.~\ref{fg1}
the test bodies are not shown to scale). The Si wafer was HF washed and then coated with
$120\pm 1~$nm of Au using an E-beam evaporator. This resulted in the rms roughness
$\delta_p=1.08~$nm measured after finishing the Casimir force measurements.
The plate was mounted on a piezoelectric tube which is used to precisely
control its position. The tube, in its turn, was mounted on a $XYZ$ linear
translational stage which is used to perform a coarse approach of the plate to
the sphere. The fine movement of the plate due to application of voltage to the
piezoelectric tube was measured by means of the second
interferometer using laser light of 520~nm wavelength. There is also a connection
to a function generator which can be used to apply different voltages to the plate
(see Fig.~\ref{fg1}).

The experimental setup was placed inside a stainless steel vacuum chamber
consisting of a mechanical scroll pump and a turbo pump connected in series to
achieve a pressure down to $10^{-9}~\mbox{Torr}\approx 1.3\times 10^{-7}~$Pa,
and an ion pump for further
reduction (see Refs.~\cite{10,15} for details). During the force measurements only
the ion pump was used in order to reduce the background mechanical noise to a minimum.
In fact ultrahigh vacuum conditions are necessary for precise measurements of the Casimir
force and are closely connected to the absence of contaminations on the Au surface.
As discussed in Sec.~I, the latter causes the electric patch effect and can lead
to a distance-dependent residue potential $V_0$ between two Au surfaces.
In addition, the desorption of contaminants from the chamber walls and their
deposition on the Au surfaces leads to a time-dependent $V_0$. To reduce the
drift rate of $V_0$, an ultra low and stable pressure is necessary. We have measured
the drift rate of $V_0$ at different chamber pressures and found that it was
0.1~mV/min at $1\times 10^{-7}~\mbox{Torr}\approx 1.3\times 10^{-5}~$Pa and less than
0.005~mV/min at $5\times 10^{-9}~\mbox{Torr}\approx 0.7\times 10^{-6}~$Pa pressure.

It has been known that to reach ultrahigh vacuum in different experiments of
surface physics the vacuum chamber is cleaned through a baking step when its temperature
is increased to more than 200\,${}^{\circ}$C to desorb all contaminants which are
then pumped out. This procedure, however, cannot be used in precise Casimir force
measurements because changes in temperature would lead to misalignment of the two
interferometers due to thermal expansion.

The removal of contamination on the Au sphere-plate surfaces used in Casimir
force measurements by means of Ar-ion bombardment was suggested in Ref.~\cite{15}.
An application of this method has helped to lower the residual potential
$V_0$ by an order of magnitude and thus, reduce the detrimental role of
electrostatic forces. However, the ions emitted by the Ar-ion gun mostly hit the
surfaces of the sphere and the plate leaving almost untouched the contaminants on the
chamber walls. As a result, after some period of time, the $V_0$ increases due to
desorption of contaminants from the chamber walls and their redeposition on the
Au surfaces of the samples.

In Ref.~\cite{58} the use of a two-step cleaning procedure in
measurements of the Casimir force was reported.
It consisted of the illumination of the entire
interior of the vacuum chamber by  UV light
followed by the Ar-ion
 bombardment of the interacting surfaces.
The UV light has long been used for removing contaminants from both the chamber
walls and surfaces of the test bodies \cite{62,63,64,65,66,67}.
UV radiation can reflect off the inner surfaces of the chamber leading to the
excellent coverage of its entire volume. The UV light can desorb water vapor and
decompose oxidative hydrocarbon from the chamber walls.

In this experiment, the UV lamp (UVB-100 Water Desorption System, RBD Instruments,
Inc.) with dimensions of $10.5^{\prime\prime}=26.67~$cm length
and $1.3125^{\prime\prime}=3.3338~$cm
diameter has been used. It was attached to the top of the vacuum chamber using a
$2.75^{\prime\prime}=6.985~$cm flange (see Fig.~\ref{fg1}). This lamp uses a hot cathode
mercury discharge tube as an emitter. It emits a combination of light with
185~nm wavelength (30\%) and with 254~nm wavelength (70\%). The radiated power
was approximately 2~W at 185~nm and 5~W at 254~nm. The UV light with 185~nm
wavelength is important because it is absorbed by oxygen and thus, leads to
the generation of ozone, whereas the UV light with the 254~nm wavelength is
absorbed by most hydrocarbons and ozone leading to their ionization and
disintegration. The resulting volatile species can then be pumped out of the
vacuum chamber.

The two-step cleaning procedure was performed as described below. At the first
step, the vacuum chamber was pumped down to the pressure of
$9\times 10^{-9}~\mbox{Torr}\approx 1.2\times 10^{-6}~$Pa by means of
the scroll mechanical pump and the turbo
pump. Next the UV lamp (see Fig.~\ref{fg1}) was turned on for 10~min.
During the UV cleaning process, the valve of the ion pump was closed to avoid
its contamination. The volatile species released by the UV light caused the
increase of chamber pressure to
$8\times 10^{-7}~\mbox{Torr}\approx 1.1\times 10^{-4}~$Pa.
These species were
pumped out by the turbo pump and mechanical pump. As a result, organic and
water contaminants on the Au surfaces of the test bodies and chamber walls were
removed leading to a modification of the residual potential $V_0$ between the
sphere and the plate.

To study this modification, a rough measurement of the $V_0$ was done for a
sphere-plate separation of $1~\mu$m before and after the UV cleaning.
We applied different voltages $V_i$ to the plate and by trial and error
found the two voltages
$V_1$ and $V_2$ which lead to the same frequency shift. Taking into account
that the frequency shift is proportional to $(V_i-V_0)^2$ (see Sec.~III),
$V_0$ was estimated as $(V_1+V_2)/2$. This results in
$V_0=49.6\pm 0.3~$mV before cleaning. During and immediately after (up to
60~min.) the UV treatment, measurements of the frequency shift were not possible
due to the fluctuating interferometer signal induced by the thermal effects of
the UV radiation.
We have found that after the
UV lamp was turned off for 60~min., and the signal was stabilized, $V_0$ reaches
a higher value in the region of 100--200~mV. The reason for this increase may be
the exposure of inorganic contaminants on the sample surface including the
possible formation of nonstable oxides of Au.

In the second step, Ar-ion-beam bombardment was used to remove any additional
organic and also inorganic contaminations, including Au oxide, from the sample
surfaces \cite{15,68,69}. For this purpose, the sphere-plate separation was
increased up to $500~\mu$m and the Ar gas from the Ar-ion gun (see Fig.~\ref{fg1})
was released into the chamber until the pressure reached the value of
$1.2\times 10^{-5}~\mbox{Torr}\approx 1.6\times 10^{-3}~$Pa
(during the Ar-ion cleaning, the ion pump remained shut off).
The Ar ions were accelerated with a 500~V potential difference.
This value was selected so that the kinetic energy of Ar ions is high enough
to break chemical bonds of Au oxide and organic molecules but low enough to avoid
any sputtering of Au surfaces. The anode current of $4~\mu$A was used as the
 ion beam flux. The filament current was 2.1~A. At these conditions, the Ar-ion
cleaning was done in several 5-min stages. After each cleaning stage the turbo
pump gate valve was opened until the pressure reached
$5\times 10^{-9}~\mbox{Torr}\approx 0.7\times 10^{-6}~$Pa
in less than 30~min. Next the value of $V_0$ was measured. This was repeated
several times until $V_0$ reached the smallest value of few millivolts.
The complete Ar-ion cleaning time was typically between 20--30~min.

To reduce mechanical noise, the turbo and mechanical pumps were valved and
then turned off and the ion pump was turned on. As a result, the two-step
cleaning procedure using the UV light and Ar ions provides us with clean
sphere-plate surfaces with low and time-stable $V_0$ ready for the force
measurements at a ultrahigh vacuum of
$5\times 10^{-9}~\mbox{Torr}\approx 0.7\times 10^{-6}~$Pa.

\section{Measurement results using small oscillation amplitude of the cantilever and
comparison with theory}

As mentioned in Sec.~I, we perform measurements of the gradient of the Casimir force
in the frequency-shift mode which is often referred to as frequency modulation.
In so doing, the cantilever with attached sphere is set to oscillate above a plate
so that the separation distance between them varies harmonically with time $t$ as
\begin{equation}
a(t)=a+A\cos\omega_r t,
\label{eq2}
\end{equation}
\noindent
where $\omega_r$ is the resonant frequency of the cantilever under the influence of
the Casimir, electric or any other force and $A$ is the oscillation amplitude.
Changes in the resonant frequency $\Delta\omega=\omega_r-\omega_0$, where
$\omega_0$ is the proper resonant frequency of the cantilever measured when it is
far away from the plate, is detected. The feedback using a phase-locked loop
(see Fig.~\ref{fg1})
allows one to keep the cantilever oscillating at its current resonant frequency with
constant amplitude \cite{10,59}.

In our experiment the sphere is subjected to the Casimir force $F$ and electric
force $F_{\rm el}$ caused by the constant voltages $V_i$ applied to the plate and
the residual potential difference $V_0$:
\begin{equation}
F_{\rm tot}(a)=F(a)+F_{\rm el}(a).
\label{eq3}
\end{equation}
\noindent
Then, in the linear regime, the frequency shift is given by \cite{10,59}
\begin{equation}
\Delta\omega=-C\frac{\partial F_{\rm tot}(a)}{\partial a},
\label{eq4}
\end{equation}
\noindent
where $C=\omega_0/(2k)$. The nonlinear corrections to this equation are investigated
in Ref.~\cite{10}. The oscillation amplitude $A$ in Eq.~(\ref{eq2}) should be chosen
from the condition that nonlinear corrections to Eq.~(\ref{eq4}) are negligibly
small.

Substituting Eq.~(\ref{eq3}) in Eq.~(\ref{eq4}) and using the exact expression for
the electrostatic force between a metallic  sphere and plate \cite{1,70}, after the
differentiation with respect to $a$ one obtains
\begin{equation}
\Delta\omega=-\gamma(V_i-V_0)^2-C\frac{\partial F(a)}{\partial a},
\label{eq5}
\end{equation}
\noindent
where  the quantity $\gamma$ is given by
\begin{eqnarray}
&&
\gamma=\frac{2\pi\epsilon_0C}{\sqrt{a(2R+a)}}
\sum_{n=1}^{\infty}{\rm csch}(n\kappa)\left\{
\vphantom{n^2}n\coth(n\kappa)\right.
\label{eq6} \\
&&~~
\left.\times
[n\coth(n\kappa)-\coth\kappa]
-{\rm csch}^2\kappa+n^2{\rm csch}^2(n\kappa)\right\}.
\nonumber
\end{eqnarray}
\noindent
Here, the parameter $\kappa$ is defined by $\cosh\kappa=1+a/R$ and
$\epsilon_0$ is the permittivity of a free space.

The first three measurement sets were taken over the separations exceeding 250~nm.
The oscillation amplitude of the cantilever in all the three sets was chosen to
be $A=10~$nm. According to Fig.~14 of Ref.~\cite{10} this ensures that the nonlinear
corrections to Eq.~(\ref{eq4}) are negligibly small. The two-step cleaning
procedure described in Sec.~II was done prior to the beginning of each
measurement set.

In each of the three sets, the measurements have been performed in the following way.
Ten different voltages $V_i$ ($i=1,\ldots\, ,10$) with the step of 0.01~V and eleven
with fixed $V_i$ ($i=11,\ldots\, ,21$) were sequentally applied to the plate and
the cantilever frequency shift was measured as a function of sphere-plate separation
at time intervals corresponding to 0.14~nm.
The frequency-shift signals at every 1~nm separation were found by interpolation
(details on the data acquisition are given in Ref.~\cite{10}).
In the first, second, and third sets, $V_i$ ($i=1,\ldots\, ,10$)
between the range (--0.04~V,\,0.06~V), (--0.049~V,\,0.051~V),  and
(--0.049~V,\,0.051~V), respectively, were applied.
In the first set,  $V_i$ ($i=11,\ldots\, ,21$)
was equal to 0.01~V, whereas in the second and third sets to 0.001~V
corresponding to the different values of $V_0$ (see below).
The relative separation between the sphere and the plate $z_{\rm rel}$ was controlled by application of voltage to the piezoelectric tube situated below the Au-coated
plate (see Fig.~\ref{fg1}). The interference fringes from the 520-nm fiber
interferometer were used to calibrate the distance moved by the plate. The absolute
sphere-plate separation was defined as $a=z_0+z_{\rm rel}$, where $z_0$ is the
separation at the closest approach determined for each of the measurement sets
separately during electrostatic calibration.

\begin{figure}[b]
\vspace*{-0cm}
\centerline{\hspace*{0.cm}
\includegraphics{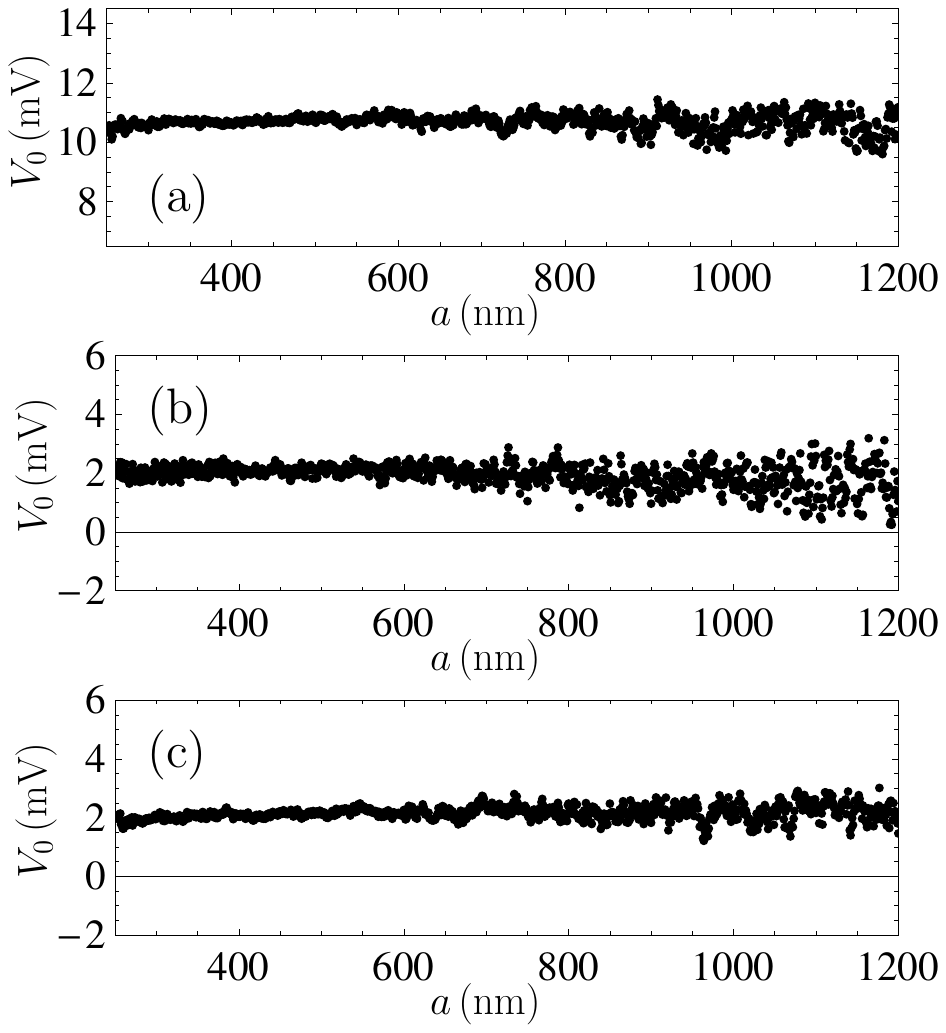}
}
\vspace*{-16.cm}
\caption{\label{fg3}
The residual potential differences between Au-coated surfaces of
 a sphere and a plate are shown by dots as functions of separation
for (a) first, (b) second, and (c) third sets of measurements
using small oscillation amplitude of the cantilever.
}
\end{figure}

The calibration of the setup, i.e., precise determination of the absolute values of
parameters $C$, $V_0$, and $z_0$, was performed with corrections
for the mechanical drift as
described in Ref.~\cite{10}. According to Eq.~(\ref{eq5}), at any separation
the frequency shift $\Delta\omega$ is described by the parabolic function of
$V_i-V_0$. By fitting the parabolas to the measured $\Delta\omega$, one finds $V_0$
from the position of the parabola maximum and $\gamma$ from the
quadratic coefficient.
The obtained values of $V_0$ over the entire measurement range from 250 to 1200~nm
with a step of 1~nm for the first, second, and third sets are shown as dots in
Figs.~\ref{fg3}(a), \ref{fg3}(b), and \ref{fg3}(c), respectively.
From Fig.~\ref{fg3} it is seen that after the two-step cleaning procedure the
residual potential difference is almost independent of separation (compared with
strongly separation-dependent $V_0$ in Fig.~2 of Ref.~\cite{15} measured between
uncleaned surfaces). The best fit of the straight line $V_0=Ka+b$  to these data leads
to the following mean values of $V_0$ and the parameters $K$ and $b$:
\begin{eqnarray}
&&
\bar{V_0}=10.7\,\mbox{mV},{\ }
K=-8.48\,\times 10^{-5}\frac{\mbox{mV}}{\mbox{nm}},{\ }
b=10.7\,\mbox{mV},
\nonumber \\
&&
\bar{V_0}=1.93\,\mbox{mV},{\ }
K=-5.33\times 10^{-4}\,\frac{\mbox{mV}}{\mbox{nm}},{\ }
b=2.32\,\mbox{mV},
\nonumber  \\
&&
\bar{V_0}=2.16\,\mbox{mV},{\ }
K=2.16\times 10^{-4}\,\frac{\mbox{mV}}{\mbox{nm}},{\ }
b=2.00\,\mbox{mV}
\nonumber
\end{eqnarray}
\noindent
for the first, second, and third measurement sets, respectively.

Next, we performed the least squares fit of the analytic expression for $\gamma$ in Eq.~(\ref{eq6})
to the value of $\gamma$ obtained from fitting the measurement data for the frequency
shift $\Delta\omega$ to the parabolas.
This was done at different separations with almost separation-independent
results for the
calibration constant $C$ and the separation at the closest approach $z_0$
(compare with Ref.~\cite{10}). The obtained mean values of these parameters are
\begin{eqnarray}
&&
z_0=248.0\pm 0.4\,\mbox{nm},{\ }C=(6.485\pm 0.006)\times 10^5\,
\frac{\mbox{s}}{\mbox{kg}},
\nonumber \\
&&
z_0=240.2\pm 0.6\,\mbox{nm},{\ }C=(6.422\pm 0.012)\times 10^5\,
\frac{\mbox{s}}{\mbox{kg}},
\nonumber  \\
&&
z_0=234.4\pm 0.5\,\mbox{nm},{\ }C=(6.529\pm 0.008)\times 10^5\,
\frac{\mbox{s}}{\mbox{kg}}
\nonumber
\end{eqnarray}
\noindent
for the first, second, and third measurement sets, respectively.
Note that the above
values for the calibration constant are almost an order of magnitude
larger than that found in Ref.~\cite{10}. This is explained by the fact that now
we use a softer cantilever with much smaller spring constant $k$.
Note also that the values of $C$ are determined independently for each set
of data, i.e., for each experiment. The small random variations in $C$ are
probably from some uncontrolled effects of the cleaning process particular
to that experiment.

For each of the three measurement sets, the 21 values of the gradient of the Casimir
force $F^{\prime}(a)=\partial F/\partial a$ at each separation distance with a step
1~nm were found from Eq.~(\ref{eq5}). The mean measured Casimir forces for each set
were obtained by averaging over 21 repetitions and their random errors were
determined at the 67\% confidence level. These random errors were added in quadrature
to the systematic errors mostly determined by the systematic error in measuring the
frequency shift (in the measurement sets 1--3 it was equal to
$5.5\times 10^{-2}~$rad/s). In this way, the total errors in each of the three
measurement sets have been obtained as functions of separation.
Then the mean gradients of the Casimir force were averaged over the three measurement
sets. The total errors of the mean force gradients $\bar{F}^{\prime}$ obtained by
averaging over the three sets is given by the mean of the total errors found for
each measurement set separately as described above \cite{71}.

\begin{figure}[b]
\vspace*{-3cm}
\centerline{\hspace*{0.5cm}
\includegraphics{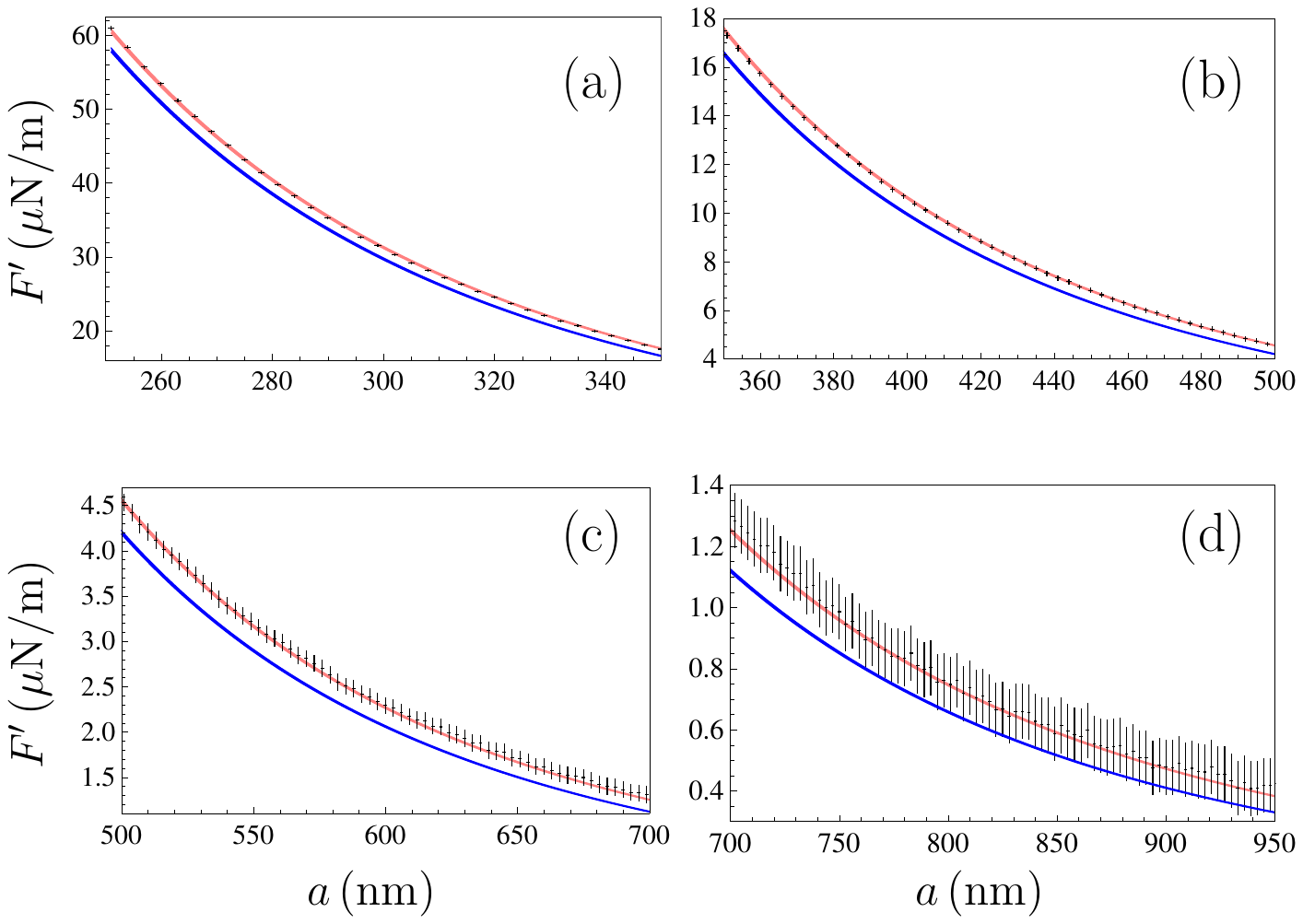}
}
\vspace*{-14.cm}
\caption{\label{fg4}
The mean gradient of the Casimir force obtained from the three
measurement sets with small oscillation amplitude of the cantilever
is shown by crosses as a function of separation within four separation
intervals. For clarity only every third experimental data point is plotted.
The bottom and top lines demonstrate theoretical predictions
of the Lifshitz theory with inclusion and neglect of the
relaxation of conduction electrons, respectively.
}
\end{figure}

The measurement results for the gradient of the Casimir force obtained from the
three sets are shown as crosses in Fig.~\ref{fg4}(a-d) over the separation range
from 250 to 950~nm (at larger separations the data are not informative).
The vertical arms of the crosses indicate the total error in measuring the force
gradient at the 67\% confidence level. The horizontal arms are determined by the
constant error in measuring the absolute separations $\Delta z=0.5~$nm.
For better visualization only each third data point is plotted in Fig.~\ref{fg4}.

We now compare experiment and theory. The thicknesses of
the Au coatings on the sphere and the plate allow one to consider these bodies
as made up entirely of Au in calculations of the Casimir force \cite{1}.
At the relatively large separations considered in this work, the surface
roughness with rms characterized in
Sec.~II leads to only negligibly small contribution which can be taken into account
perturbatively. For a small ratio $a/R<0.022$ the calculation of
the gradient of the Casimir
force can be performed within the proximity force approximation with inclusion
of the first-order corrections to this approximation in $a/R$ \cite{53a,52,53,54,55}
computed with inclusion of the real material properties. As a result, the gradient
of the Casimir force acting between a sphere and a plate is given by
\begin{equation}
F_{\rm theor}^{\prime}(a)=-2\pi R\left[1+\beta(a,R)\frac{a}{R}\right]
\left(1+10\frac{\delta_s^2+\delta_p^2}{a^2}\right)P(a),
\label{eq9}
\end{equation}
\noindent
where we use numerical values for the function $\beta$ computed in Ref.~\cite{55}
using the extrapolation of the optical data for Au to low frequencies by means of the
Drude and plasma models,  and $P(a)$ is the Casimir pressure between two Au plates
computed at the temperature $T=20~{}^{\circ}$C of the experiment.
This pressure is expressed by the commonly known Lifshitz formula \cite{1,2,3,4,5}
\begin{equation}
P(a)=-\frac{k_BT}{\pi}
\sum_{l=0}^{\infty}{\vphantom{\sum}}^{\prime}\!\!\int_{0}^{\infty}
\!\!\!\!q_lk_{\bot}dk_{\bot}\!\sum_{\alpha}
\frac{1}{r_{\alpha}^{-2}(i\xi_l,k_{\bot})e^{2aq_l}-1}.
\label{eq10}
\end{equation}
\noindent
Here, $q_l=(k_{\bot}^2+\xi_l^2/c^2)^{1/2}$, the integration is performed with respect
to the magnitude of the projection of the wave vector on the plane of plates,
$\xi_l=2\pi k_BTl/\hbar$ with $k_B$ being the Boltzmann constant,
$l=0,\,1,\,2,\,\ldots$ are the Matsubara frequencies, the summation in $\alpha$ is
made over the two independent polarizations of the electromagnetic field, transverse
magnetic ($\alpha={\rm TM}$) and transverse electric ($\alpha={\rm TE}$),
and the reflection coefficients are expressed as
\begin{equation}
r_{\rm TM}(i\xi_l,k_{\bot})=\frac{\varepsilon_lq_l-k_l}{\varepsilon_lq_l+k_l},
\quad
r_{\rm TE}(i\xi_l,k_{\bot})=\frac{q_l-k_l}{q_l+k_l},
\label{eq11}
\end{equation}
\noindent
where the dielectric permittivity of Au is taken at the pure imaginary Matsubara
frequencies $\varepsilon_l=\varepsilon(i\xi_l)$ and
\begin{equation}
k_l=\left(k_{\bot}^2+\varepsilon_l\frac{\xi_l^2}{c^2}\right)^{1/2}.
\label{eq12}
\end{equation}

Numerical computations of the gradient of the Casimir force have been performed
by Eqs.~(\ref{eq9})--(\ref{eq12}) in the framework of the two approaches discussed
in Sec.~I, i.e., by using $\varepsilon_l$ obtained from the optical data of Au
\cite{72} extrapolated down to zero frequency by means of either the dissipative Drude
or the dissipationless plasma models
\begin{equation}
\varepsilon_D(i\xi)=1+\frac{\omega_p^2}{\xi(\xi+\tau^{-1})}, \quad
\varepsilon_p(i\xi)=1+\frac{\omega_p^2}{\xi^2},
\label{eq13}
\end{equation}
\noindent
where $\hbar\omega_p=9.0~$eV and $\hbar\tau^{-1}=35~$meV are the energies corresponding
to the plasma frequency and relaxation parameter $\gamma=\tau^{-1}$
($\tau$ is the relaxation time) \cite{72}.

The computational results are presented in Fig.~\ref{fg4}(a--d) as a function of
separation by the bottom and top lines obtained using the Drude and the plasma
model approaches, respectively. The width of the lines characterizes the size
of the theoretical error which is largerly determined by inaccuracies in the
optical data of Au.

{}From Fig.~\ref{fg4} one can conclude that the theoretical predictions using the
Drude model approach (i.e., taking into account the energy losses by conduction
electrons) are excluded by the data over the separation range from 250 to 850~nm.
As to the plasma model approach, which disregards the energy losses of conduction
electrons, it is consistent with the measurement data over the entire  separation
region. Similar results have been obtained previously in the separation range
up to 420~nm with the help of the dynamic AFM \cite{10,15} and in the separation
range up to 750~nm with the help of micromechanical torsional oscillator
\cite{1,2,6,7,8,9,14}. In Ref.~\cite{58} only one of the three measurement sets,
presented in this paper, allowed an exclusion of the Drude model approach in the
region of separations from 250 to 820~nm.

\begin{figure}[b]
\vspace*{-0cm}
\centerline{\hspace*{0.3cm}
\includegraphics{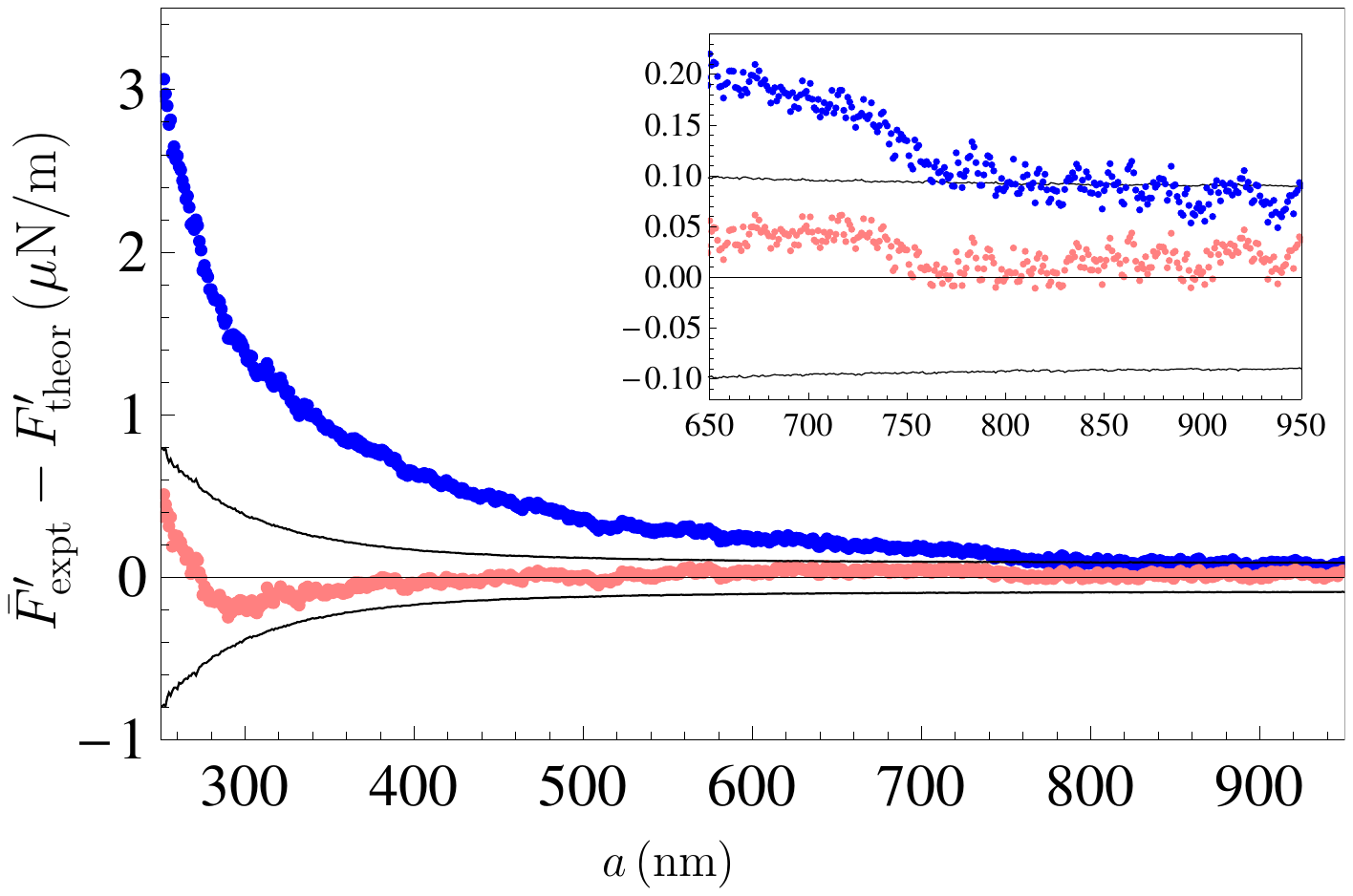}
}
\vspace*{-17.cm}
\caption{\label{fg5}
The differences between the mean gradient of the Casimir force
obtained from the three measurement sets with small oscillation
amplitude of the cantilever and theoretical gradients calculated
with inclusion and neglect of the
relaxation of conduction electrons are shown as functions of
separations by the top and bottom sets of dots, respectively.
The two lines are formed by the boundary points of the confidence
intervals for the quantity
$\bar{F}_{\rm expt}^{\prime}-{F}_{\rm theor}^{\prime}$.
In the inset the region of separations from 650 to 950~nm is shown on
an enlarged scale.
}
\end{figure}

The obtained results are confirmed using another method of comparison
between experiment and theory which considers the differences between mean
experimental $\bar{F}_{\rm expt}^{\prime}$ and theoretical
${F}_{\rm theor}^{\prime}$ gradients of the
Casimir force. These differences are plotted as
dots in Fig.~\ref{fg5} with a step of 1~nm by the top and bottom sets found using
the Drude and the plasma model approaches for ${F}_{\rm theor}^{\prime}$,
respectively. In doing so the experimental gradients are the mean values
obtained from the three measurement sets. The lower and upper solid lines
in Fig.~\ref{fg5} are formed by the smoothly joined boundary points of the
confidence intervals for the differences
$\bar{F}_{\rm expt}^{\prime}-{F}_{\rm theor}^{\prime}$.
The width of these intervals is equal to twice the total error in the quantity
$\bar{F}_{\rm expt}^{\prime}-{F}_{\rm theor}^{\prime}$.
The latter is found by combining in quadrature the already known
total experimental error of
$\bar{F}_{\rm expt}^{\prime}$ and the total theoretical error of
${F}_{\rm theor}^{\prime}$, which is determined by the errors arising from the
inaccuracy of the optical data and from the calculation of the force gradient at
the separation distance determined with an error $\Delta z$ (see Refs.~\cite{1,7}
for details). In the inset, the region of separations from 650 to 950~nm is
shown on an enlarged scale.

The meaning of the confidence band in between the solid lines is the following.
If the theoretical approach is consistent with the data within some separation
interval at the 67\% confidence level, no less than 67\% of the dots in this
interval should belong to the confidence band. On the other hand, the theoretical
approach is excluded by the data within some interval at the same confidence
level if more than 33\% of the dots fall outside the confidence band \cite{1,7,73}.
From Fig.~\ref{fg5} one can see that the plasma model approach is consistent with the
data over the entire measurement range from 250 to 950~nm. At the same time, the
Drude model approach is excluded by the data at all separations below 850~nm
(note that although some dots of the top set belong to the confidence band at
separations from 770 to 850~nm, the number of these dots does not reach 67\%
of all the dots belonging to this interval).

Thus, the consideration of three measurement sets allows us not only to confirm
the results obtained in Ref.~\cite{58} from a single set, but also to increase
the upper boundary of the separation interval up to 850~nm, where the Drude
model approach is
excluded by the data.

\section{Comparison with theory of the measurement results with larger
oscillation amplitude}

Next we have performed one more set of measurements of the gradient of the
Casimir force with much larger separation  of the closest approach between the
sphere and the plate. This allows the use of a larger oscillation amplitude of
the cantilever $A\approx 20$~nm while preserving the linearity of Eq.~(\ref{eq4}).

After finishing the third measurement set it was checked
to confirm that the vacuum conditions
in the chamber remain stable. Because of this, it was not necessary to repeat
the two-step cleaning procedure done previously before each of the first three measurement sets.
Measurements have been performed in the separation region from 600~nm to 2~$\mu$m in
a similar way to the first three sets but the values of the applied voltages
have been changed for larger ones. Ten different voltages  $V_i$ $(i=1,\ldots,10)$ with
a step of 0.01$V$ varied in the interval (--0.092~V, 0.108~V) whereas eleven fixed
voltages $V_i$ $(i=11,\ldots,21)$ were equal to 0.008~V close to $V_0$.

\begin{figure}[b]
\vspace*{-10cm}
\centerline{\hspace*{3.cm}
\includegraphics{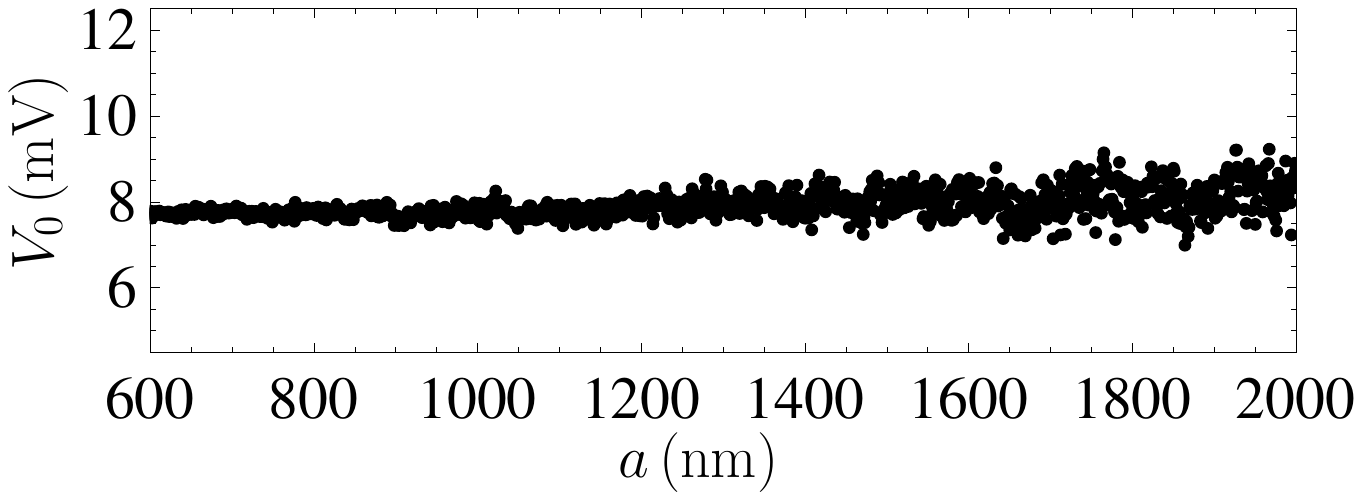}
}
\vspace*{-10.cm}
\caption{\label{fg6}
The residual potential difference between Au-coated surfaces of
 a sphere and a plate is shown by dots as a function of separation
 for the set of measurements with larger oscillation amplitude
of the cantilever.
}
\end{figure}

The calibration of the setup was performed as described in Sec.~III. First the
residual potential difference $V_0$ was determined at each separation with a step of 1~nm. The obtained results are shown by the dots in Fig.~\ref{fg6} as a function of separation. Similar to Fig.~\ref{fg3}, the residual potential difference is almost
separation-independent. This confirms that the surfaces of both test bodies are
 sufficiently clean and ready for the force measurements. The best fit of the straight
line $V_0=Ka+b$ to these data results in
$K=3.23\times 10^{-4}~$mV/nm and $b=7.50~$mV in close analogy to the results
obtained in the first three measurement sets (see Sec.~III).
The mean value of the $V_0$ in this case was found to be $\bar{V}_0=7.92~$mV.

Next the separation at the closest approach $z_0=571.9\pm 1.1~$nm and the calibration
constant $C=(6.342\pm 0.004)\times 10^5~$s/kg were determined from the fit as
described above. The latter is again in agreement with the values obtained in the
first three measurement sets.

Then the 21 values of the gradient of the Casimir force at each separation with a
step of 1~nm were calculated using Eq.~(\ref{eq5}). These values were averaged and
the random error was found at the 67\% confidence level as a function of separation.
The systematic error in measurements of the frequency shift for this set was equal
to $4.0\times 10^{-2}~$rad/s. The decrease of this error as compared to the value
used in Sec.~III is due to the fact that with the larger amplitude of the cantilever
oscillations the corresponding interferometer signal is increased. This increases
the signal-to-noise ratio leading to a reduced error in the determination of the
cantilever frequency shift from the sphere-plate interaction forces.
This error is used in the calculation of
the systematic error in measurements of the force gradient
by Eq.~(\ref{eq4}) and combined in quadrature with the random error to obtain
the total experimental error in the gradient of the Casimir force $\bar{F}^{\prime}$.

\begin{figure}[b]
\vspace*{-1cm}
\centerline{\hspace*{0.cm}
\includegraphics{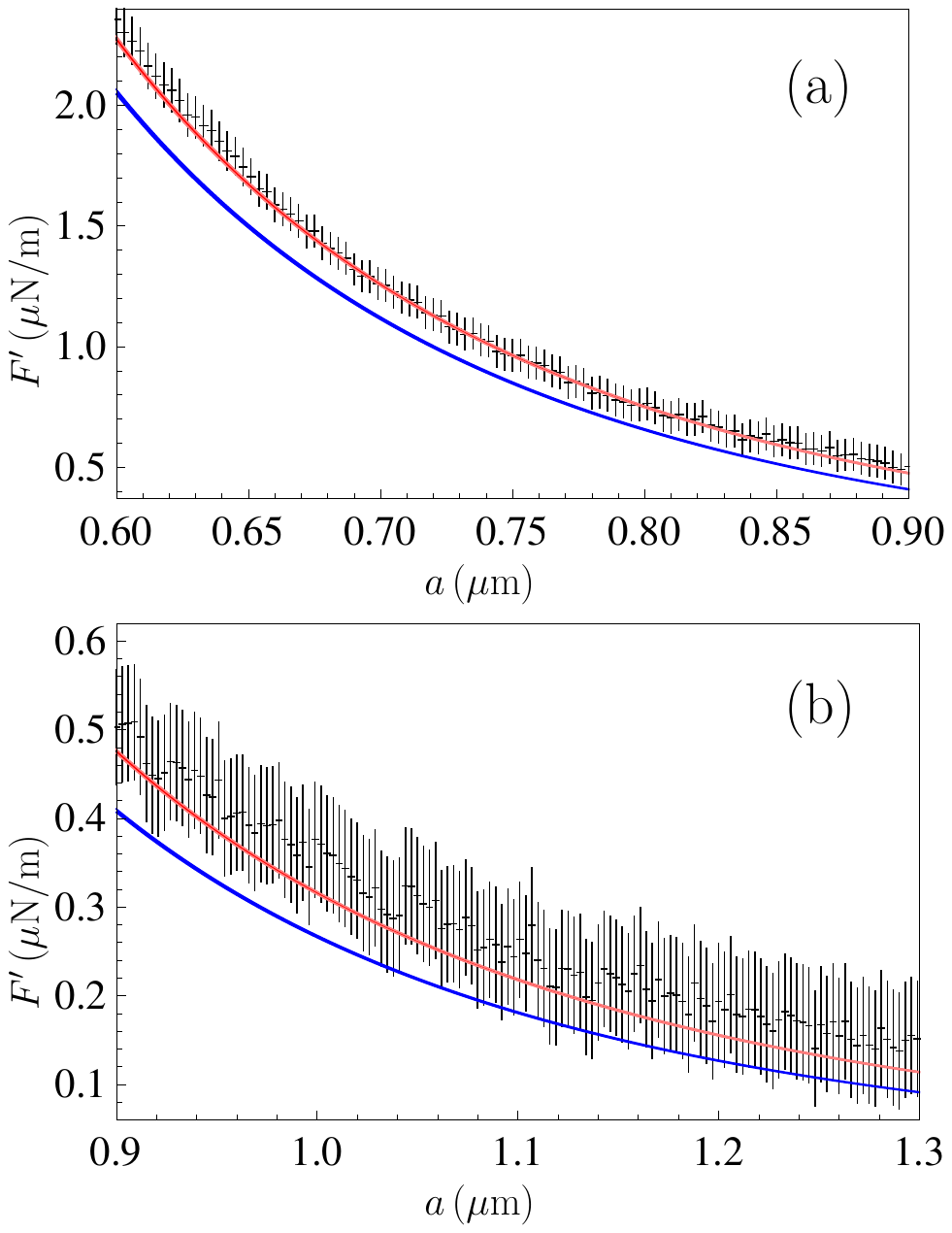}
}
\vspace*{-13.cm}
\caption{\label{fg7}
The mean gradient of the Casimir force obtained from the
measurement set with larger oscillation amplitude of the cantilever
is shown as a function of separation within two separation intervals.
For clarity only every third experimental data point is plotted.
The bottom and top lines demonstrate theoretical predictions
of the Lifshitz theory with inclusion and neglect of the
relaxation of conduction electrons, respectively.
}
\end{figure}

The measured gradients are shown as crosses in Fig.~\ref{fg7}(a,b) over the separation
region from 600~nm to 1.3~$\mu$m.
The arms of the crosses indicate the total error in measuring the force
gradient and in measuring the absolute separations $\Delta z=1.1~$nm.
Only each third cross is plotted in Fig.~\ref{fg7} to make the figure more informative.

The theoretical force gradients are computed as described in Sec.~III.
The computational results obtained using the Lifshitz theory and the optical data for
Au extrapolated down to zero frequency by means of either the Drude of the plasma
models are shown in Fig.~\ref{fg7} as functions of separation by the bottom and top
lines, respectively. As is seen in this figure, the theoretical predictions using the
dissipationless plasma model are consistent with the measurement data over the entire range
from 600~nm to $1.3~\mu$m. From the same figure, we can conservatively conclude that
the predictions of the Lifshitz theory using the dissipative Drude model for extrapolation
are excluded at all separations up to $1.1~\mu$m. Thus, the range of separations
where the Drude model is excluded by the data has been significantly extended.

\begin{figure}[b]
\vspace*{-0cm}
\centerline{\hspace*{0.3cm}
\includegraphics{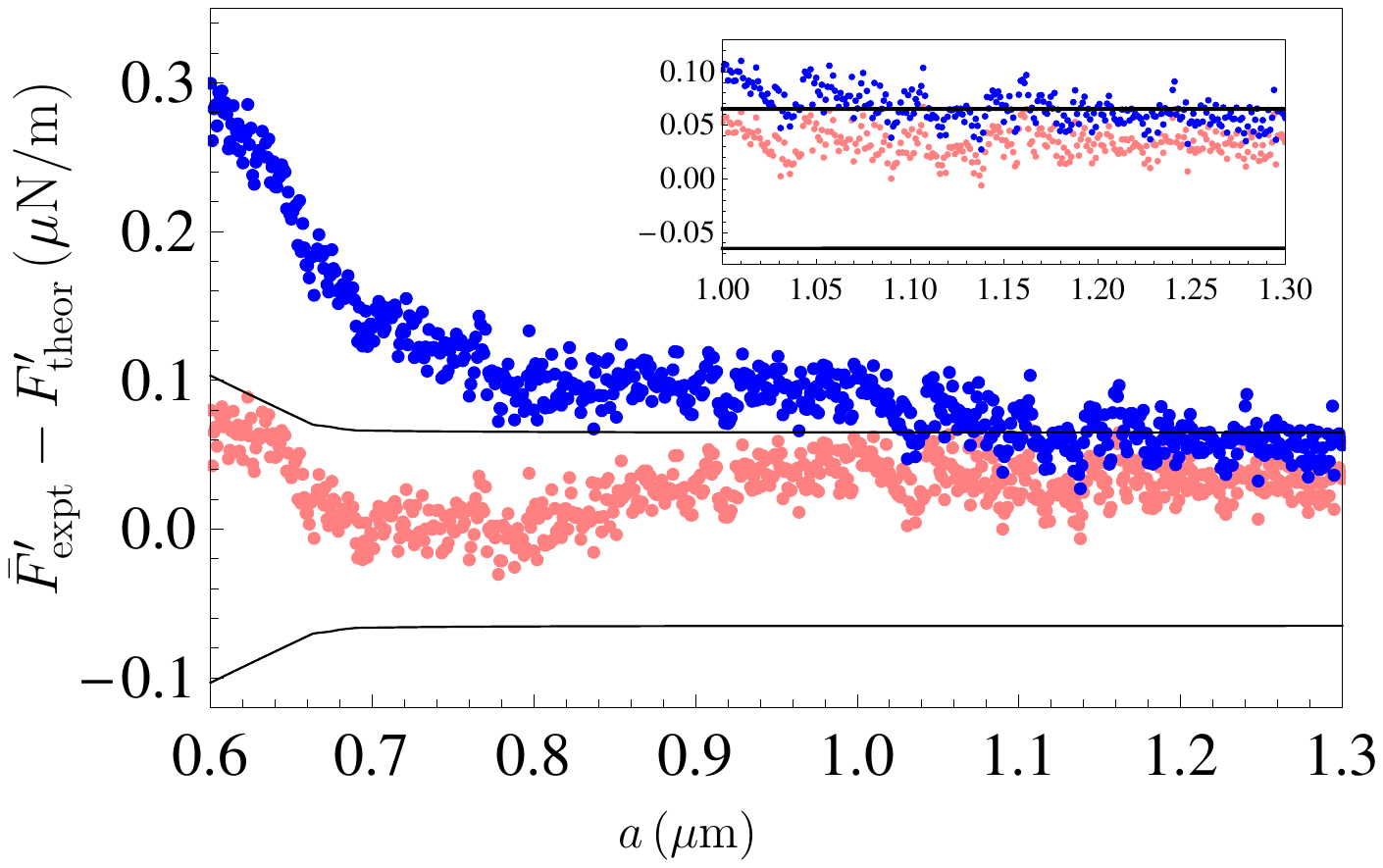}
}
\vspace*{-17.cm}
\caption{\label{fg8}
The differences between the mean gradient of the Casimir force
obtained from the measurement set with larger oscillation
amplitude of the cantilever and theoretical gradients calculated
with inclusion and neglect of the
relaxation of conduction electrons are shown as functions of
separations by the top and bottom sets of dots, respectively.
The two lines are formed by the boundary points of the confidence
intervals for the quantity
$\bar{F}_{\rm expt}^{\prime}-{F}_{\rm theor}^{\prime}$.
In the inset the region of separations from 1 to $1.3~\mu$m is shown on
an enlarged scale.
}
\end{figure}

Now we compare with theory the measurement data obtained at larger separations with
increased oscillation amplitude using another statistical approach discussed
in Sec.~III. The differences between $\bar{F}_{\rm expt}^{\prime}$ and
${F}_{\rm theor}^{\prime}$ are plotted in Fig.~\ref{fg8} as dots with a step of 1~nm.
The top and bottom sets of dots are obtained using ${F}_{\rm theor}^{\prime}$
calculated using the Drude and plasma model approaches, respectively.
The two solid lines are formed by the boundary points of the confidence intervals for
$\bar{F}_{\rm expt}^{\prime}-{F}_{\rm theor}^{\prime}$
determined as discussed in Sec.~III. In the inset, the region of largest separations
from 1 to $1.3~\mu$m is shown on an enlarged scale to gain a better understanding.

{}From Fig.~\ref{fg8} it is seen that all the points of the bottom set belong to
the confidence interval, i.e., the plasma model approach is consistent with the data.
The Lifshitz theory combined with the Drude model approach is excluded by the data
over the region of separations from 0.6 to $1.1~\mu$m. In the intervals belonging
to this range more than 33\% of dots lie outside the confidence band.
Therefore, the second method of comparison between experiment and theory
leads to the same conclusions as the first one which means that in this experiment
the region of separations where the Drude model approach is excluded is
extended up to $1.1~\mu$m.

\section{Conclusions and discussion}

In the foregoing, we have presented a description of the experiment on measuring
the gradient of the Casimir force between metallic surfaces of a sphere and a plate
cleaned by means of a two-step cleaning procedure using the UV light and Ar-ion
bombardment. Compared to Ref.~\cite{58}, two additional measurement sets within
the same separation range are reported here, as well as the measurement results
at larger separations with a factor of two larger amplitude of the
cantilever oscillations. The latter allowed one to significantly increase the range
of separations where the experiment discriminates between the two theoretical
approaches used in the literature on Casimir physics. Specifically, theoretical
predictions based on the Lifshitz theory in combination with the
dissipative Drude model for
conduction electrons were excluded by the measurement data up to the separation
of $1.1~\mu$m (to compare with 820~nm in Ref.~\cite{58}).

As discussed in Sec.~I, both the Casimir puzzle for metallic test bodies and the
Casimir conundrum for dielectric and semiconductor ones are the problems which
still remain to be solved. Prior to this there already was a reasonably good picture
of the experimental situation concerning the Casimir puzzle at separations below
a few hundred nanometers, but the situation at separations above $0.8~\mu$m
remained completely unresolved. This made harder the theoretical solution to the
problem and called for precise measurements of the Casimir interaction in the
micrometer separation range. Several experiments of this kind directed to the
resolution of the Casimir puzzle and Casimir conundrum have been proposed
recently \cite{74,75,76,77,78}.

The main improvements made in the experiment presented here are the use of much
softer cantilever, which allowed the increase of the calibration constant by up to
an order of magnitude, and implementation of the two-step cleaning procedure
by means of the UV light and Ar ions, which resulted in ultra clean surfaces of
both the internal walls of vacuum chamber and of the test bodies at ultrahigh
($<5\times 10^{-9}~\mbox{Torr}\approx 0.7\times 10^{-6}~$Pa) and stable vacuum. This allowed
to reach  low (a few mV) and
stable residual potential difference
which was independent of separation for the regular
(not specially selected) samples. The introduction of these tools has made it
possible to discriminate between the theoretical predictions with inclusion
and neglect of the dissipative
relaxation of conduction electrons in three measurement sets with
oscillation amplitude of cantilever equal to 10~nm up to larger separation
distances. As a result, the Drude model approach was excluded by the data over the
separation range from 250 to 850~nm, and the plasma model approach was found to be
consistent with the data.

It was checked that in the separation range from 0.6 to $2~\mu$m the oscillator used
is still in the linear regime when the oscillation amplitude is increased to 20~nm.
With this increased oscillation amplitude, the measurements of the gradient of the
Casimir force have been repeated and compared with the same two theoretical
approaches. It is found that the plasma model approach neglecting the relaxation
of conduction electrons is again consistent with the data over the entire measurement
range from 0.6 to $1.3~\mu$m. The Drude model approach taking into account the
relaxation properties of conduction electrons was excluded over the separation range
from 0.6 to $1.1~\mu$m. Thus, by the results of measurements with 10- and 20-nm
oscillation amplitudes of the cantilever the Drude model approach is excluded by the
data over the separation range from 250 to 1100~nm.

The problem of why the Lifshitz theory is in contradiction with the measurement data
when it takes into account the relaxation properties of conduction electrons
at low frequencies is discussed in the literature but there is yet no consensus on how
this puzzle can be explained. In Refs.~\cite{58,78,79} it was hypothesized that
a material system might not respond similarly to electromagnetic fields with nonzero
field strength and to fluctuations with zero field strength but nonzero dispersion.
This hypothesis does not necessarily assume a violation of the
fluctuation-dissipation theorem, but might be connected with the phenomenological
character of the Drude model which describes well the response of metals
to real electromagnetic fields on the mass-shell but fails to give an adequate
description for fluctuations that are not on the mass-shell.
Future investigations will shed new light on this problem.

\section*{Acknowledgments}
The work of M.L., J.X.~and U.M.~was partially supported by the NSF
grant PHY-1607749.
M.L., J.X.~and U.M.~acknowledge discussions with R.\ Schafer and Tianbai Li.
G.L.K.\ and V.M.M.\ were partially supported by the Peter the Great
Saint Petersburg Polytechnic University in the framework of the
Program ``5--100--2-20".
V.M.M.~was partially funded by the Russian Foundation for Basic
Research, Grant No. 19-02-00453 A. His work was also partially
supported by the Russian Government Program of Competitive Growth
of Kazan Federal University.

\newpage

\begin{thebibliography}{00}
\bibitem{1}
M.~Bordag, G.~L.~Klimchitskaya, U.\ Mohideen, and
V.\ M.\ Mostepanenko, {\it Advances in the Casimir Effect}
(Oxford University Press, Oxford, 2015).
\bibitem {2}
G.~L.~Klimchitskaya, U. Mohideen, and V.\ M.\ Mostepanenko,
The Casimir force between real materials: Experiment and theory,
 Rev. Mod. Phys. {\bf 81}, 1827 (2009).
\bibitem{3}
A.~W.~Rodrigues, F.~Capasso, and S.~G.~Johnson,
The Casimir effect in microstructured geometries,
Nat. Photon. {\bf 5}, 211 (2011).
\bibitem{4}
L.~M.~Woods, D.~A.~R.~Dalvit, A.~Tkatchenko, P.\ Rodriguez-Lopez,
A.\ W.\ Rodriguez, and R.\ Podgornik,
Materials perspective on Casimir and van der Waals interactions,
Rev. Mod. Phys. {\bf 88}, 045003 (2016).
\bibitem{5}
E.~M.~Lifshitz,
The theory of molecular attractive forces between solids,
Zh. Eksp. Teor. Fiz. {\bf 29}, 94 (1955)
[Sov. Phys. JETP  {\bf 2}, 73 (1956)].
\bibitem {5a}
G.~L.~Klimchitskaya and V.\ M.\ Mostepanenko,
Experiment and theory in the Casimir effect,
Contemp. Phys. {\bf 47}, 131 (2006).
\bibitem {5b}
G.~L.~Klimchitskaya and V.\ M.\ Mostepanenko,
Graphene may help to solve the Casimir conundrum in indium tin oxide
systems,
Phys. Rev. B {\bf 98}, 035307 (2017).
\bibitem{6}
R.~S.~Decca, E.~Fischbach, G.~L.~Klimchitskaya,
D.~E.~Krause, D.~L\'opez, and V.~M.~Mostepanenko,
Improved tests of extra-dimensional physics and thermal
quantum field theory from new Casimir force measurements,
Phys. Rev. D {\bf 68}, 116003 (2003).
\bibitem{7}
R.{\ }S. Decca, D. L\'opez, E. Fischbach, G.{\ }L. Klimchitskaya,
D.{\ }E. Krause, and V.{\ }M.\ Mostepanenko,
Precise comparison of theory and new experiment for the
Casimir force leads to stronger constraints on thermal
quantum effects and long-range interactions,
Ann. Phys. (N.Y.) {\bf 318},
37 (2005).
\bibitem{8}
R.~S.~Decca, D.~L\'opez, E.~Fischbach, G.~L.~Klimchitskaya,
D.~E.~Krause, and V.~M.~Mostepanenko,
Tests of new physics from precise measurements of the Casimir
pressure between two gold-coated plates,
Phys. Rev. D {\bf 75}, 077101 (2007).
\bibitem{9}
R.~S.~Decca, D.~L\'opez, E.~Fischbach, G.~L.~Klimchitskaya,
D.~E.~Krause, and V.~M.~Mostepanenko,
Novel constraints on light elementary particles and
extra-dimensional physics from the Casimir effect,
Eur. Phys. J. C {\bf 51}, 963 (2007).
\bibitem{10}
C.-C.~Chang, A.~A.~Banishev, R.~Castillo-Garza,
G.~L.~Klimchitskaya, V.\ M.\ Mostepanenko, and U.\ Mohideen,
Gradient of the Casimir force between Au surfaces of
a sphere and a plate measured using an atomic force microscope
in a frequency-shift technique,
Phys. Rev. B {\bf 85}, 165443 (2012).
\bibitem{11}
A.~A.~Banishev, C.-C.~Chang,
G.~L.~Klimchitskaya, V.\ M.\ Mostepanenko, and U.\ Mohideen,
Measurement of the gradient of the Casimir force between
a nonmagnetic gold sphere and a magnetic nickel plate,
Phys. Rev. B {\bf 85}, 195422 (2012).
\bibitem{12}
A.~A.~Banishev,
G.~L.~Klimchitskaya, V.\ M.\ Mostepanenko, and U.\ Mohideen,
Demonstration of the Casimir Force between Ferromagnetic
Surfaces of a Ni-Coated Sphere and a Ni-Coated Plate,
Phys. Rev. Lett. {\bf 110}, 137401 (2013).
\bibitem{13}
A.~A.~Banishev,
G.~L.~Klimchitskaya, V.\ M.\ Mostepanenko, and U.\ Mohideen,
Casimir interaction between two magnetic metals in comparison
with nonmagnetic test bodies,
Phys. Rev. B {\bf 88}, 155410 (2013).
\bibitem{14}
G.~Bimonte, D.~L\'{o}pez, and R.{\ }S.\ Decca,
Isoelectronic determination of the thermal Casimir force,
Phys. Rev. B {\bf 93}, 184434 (2016).
\bibitem{15}
J.~Xu,
G.~L.~Klimchitskaya, V.\ M.\ Mostepanenko, and U.\ Mohideen,
Reducing detrimental electrostatic effects in Casimir-force measurements
and Casimir-force-based microdevices,
Phys. Rev. A {\bf 97}, 032501 (2018).
\bibitem{16}
V.{\ }B.\ Bezerra, G.{\ }L.\ Klimchitskaya, and V.{\ }M.\ Mostepanenko,
Thermodynamical aspects of the Casimir
force between real metals at nonzero temperature,
Phys. Rev. A {\bf 65}, 052113 (2002).
\bibitem{17}
V.{\ }B.\ Bezerra, G.{\ }L.\ Klimchitskaya, and V.{\ }M.\ Mostepanenko,
Correlation of energy and free energy for the thermal Casimir
force between real metals,
Phys. Rev. A {\bf 66}, 062112 (2002).
\bibitem{18}
V.{\ }B.\ Bezerra, G.{\ }L.\ Klimchitskaya, V.{\ }M.\ Mostepanenko,
and C.\ Romero,
Violation of the Nernst heat theorem in the theory of
the thermal Casimir force between Drude metals,
Phys. Rev. A {\bf 69}, 022119 (2004).
\bibitem{19}
M.~Bordag and I.~G.~Pirozhenko,
Casimir entropy for a ball in front of a plane,
Phys. Rev. D {\bf 82}, 125016 (2010).
\bibitem{20}
G.~L.~Klimchitskaya
and V.~M.~Mostepanenko,
Low-temperature behavior of the Casimir free energy and entropy
of metallic films,
{Phys. Rev.} A {\bf 95}, 012130 (2017).
\bibitem{21}
G.~L.~Klimchitskaya and C.~C.~Korikov,
Analytic results for the Casimir free energy between ferromagnetic metals,
Phys. Rev. A {\bf 91}, 032119 (2015).
\bibitem{23}
F. Chen, G. L. Klimchitskaya, V. M. Mostepanenko, and U. Mohideen,
Demonstration of optically modulated dispersion forces,
Opt. Express {\bf 15}, 4823 (2007).
\bibitem{24}
F. Chen, G. L. Klimchitskaya, V. M. Mostepanenko, and U. Mohideen,
Control of the Casimir force by the modification of dielectric properties with light, Phys. Rev. B {\bf 76}, 035338 (2007).
\bibitem{25}
C.-C.~Chang, A.~A.~Banishev,
G.~L.~Klimchitskaya, V.\ M.\ Mostepanenko, and U.\ Mohideen,
Reduction of the Casimir Force from Indium Tin Oxide Film by UV Treatment,
Phys. Rev. Lett. {\bf 107}, 090403 (2011).
\bibitem{26}
A.~A.~Banishev, C.-C.~Chang, R.\ Castillo-Garza,
G.~L.~Klimchitskaya, V.\ M.\ Mostepanenko, and U.\ Mohideen,
Modifying the Casimir force between indium tin oxide film
and Au sphere,
Phys. Rev. B {\bf 85}, 045436 (2012).
\bibitem{27}
G. L. Klimchitskaya and V. M. Mostepanenko, Conductivity of dielectric and thermal atom-wall interaction, J. Phys. A: Math. Theor. {\bf 41}, 312002 (2008).
\bibitem{22}
J.~M.~Obrecht, R.~J.~Wild, M.~Antezza, L.~P.~Pitaevskii,
S.~Stringari, and E.~A.~Cornell,
Measurement of the Temperature Dependence of the Casimir-Polder
Force,
Phys. Rev. Lett. {\bf 98}, 063201 (2007).
\bibitem{28}
B. Geyer, G. L. Klimchitskaya, and V. M. Mostepanenko, Thermal quantum field theory and the Casimir interaction between dielectrics,
    Phys. Rev. D {\bf 72}, 085009 (2005).
\bibitem{29}
G. L. Klimchitskaya and C. C. Korikov,
Casimir entropy for magnetodielectrics,
J. Phys.: Condens. Matter {\bf 27}, 214007 (2015).
\bibitem{30}
G. L. Klimchitskaya and V. M. Mostepanenko,
Casimir free energy of dielectric films: classical limit, low-temperature behavior
and control, J. Phys.: Condens. Matter {\bf 29}, 275701 (2017).
\bibitem{31}
Y.~Srivastava, A.~Widom, and M.~H.~Friedman,
Microchips as Precision Quantum-Electrodynamic Probes,
Phys. Rev. Lett. {\bf 55}, 2246 (1985).
\bibitem{32}
H. B. Chan, V. A. Aksyuk, R. N. Kleiman, D. J. Bishop, and F. Capasso,
Quantum Mechanical Actuation of Microelectromechanical Systems by the Casimir Force, Science {\bf 291}, 1941 (2001).
\bibitem{33}
H. B. Chan, V. A. Aksyuk, R. N. Kleiman, D. J. Bishop, and F. Capasso, Nonlinear Micromechanical Casimir Oscillator,
    Phys. Rev. Lett. {\bf 87}, 211801 (2001).
\bibitem{34}
E.~Buks and M.~L.~Roukes,
Stiction, adhesion, and the Casimir effect
in micromechanical systems,
Phys. Rev. B {\bf 63}, 033402 (2001).
\bibitem{35}
E.~Buks and M.~L.~Roukes,
Metastability and the Casimir effect in
micromechanical systems,
Europhys. Lett. {\bf 54}, 220 (2001).
\bibitem{36}
J. B\'{a}rcenas, L. Reyes, and R. Esquivel-Sirvent, Scaling of micro- and
nanodevices actuated by the Casimir force,
Appl. Phys. Lett. {\bf 87}, 263106 (2005).
\bibitem {37}
G.\ Palasantzas,
Contact angle influence on the pull-in voltage of
microswitches in the presence of capillary and quantum
vacuum effects,
J. Appl. Phys. {\bf 101}, 053512 (2007).
\bibitem {38}
G.\ Palasantzas,
Pull-in voltage of microswitch rough plates
in the presence of electromagnetic and acoustic Casimir forces,
J. Appl. Phys. {\bf 101}, 063548 (2007).
\bibitem{39}
R. Esquivel-Sirvent and R. P\'{e}rez-Pascual, Geometry and charge carrier induced stability in Casimir actuated nanodevices,
    Eur. Phys. J. B {\bf 86}, 467 (2013).
\bibitem{40}
W. Broer, G. Palasantzas, J. Knoester, and V. B. Svetovoy, Significance of the Casimir force and surface roughness for actuation dynamics of MEMS, Phys. Rev. B {\bf 87}, 125413 (2013).
\bibitem{41}
M. Sedighi, W. H. Broer, G. Palasantzas, and B. J. Kooi, Sensitivity of micromechanical actuation on amorphous to crystalline phase transformations under the influence of Casimir forces, Phys. Rev. B {\bf 88}, 165423 (2013).
\bibitem{42}
J. Zou, Z. Marcet, A. W. Rodriguez, M. T. H. Reid, A. P. McCauley, I. I. Kravchenko, T. Lu, Y. Bao, S. G. Johnson, and H. B. Chan, Casimir forces on a silicon micromechanical chip, Nat. Commun. {\bf 4}, 1845 (2013).
\bibitem{43}
W. Broer, H. Waalkens, V. B. Svetovoy, J. Knoester, and G. Palasantzas,
Nonlinear Actuation Dynamics of Driven Casimir Oscillators with Rough Surfaces,
Phys. Rev. Appl. {\bf 4}, 054016 (2015).
\bibitem{44}
N. Inui, Optical switching of a graphene mechanical switch using the Casimir effect,
J. Appl. Phys. {\bf 122}, 104501 (2017).
\bibitem{45}
G. L. Klimchitskaya, V. M. Mostepanenko, V. M. Petrov, and T. Tschudi,
Optical Chopper Driven by the Casimir Force, Phys. Rev. Applied {\bf 10}, 014010 (2018).
\bibitem{46}
F. Tajik, M. Sedighi, A. A. Masoudi, H. Waalkens, and G. Palasantzas,
Sensitivity of chaotic behavior to low optical frequencies of a
double-beam torsional actuator,
Phys. Rev. E {\bf 100}, 012201 (2019).
\bibitem{47}
M.~Bordag,
G.~L.~Klimchitskaya, and V.\ M.\ Mostepanenko,
The Casimir force between plates with small deviations from plane
parallel geometry,
Int. J. Mod. Phys.  A {\bf 10}, 2661 (1995).
\bibitem{48}
W. Broer,  G. Palasantzas, J. Knoester, and V. B. Svetovoy,
Roughness correction to the Casimir force at short separations:
Contact distance and extreme value statistics,
Phys. Rev. B {\bf 85}, 155410 (2012).
\bibitem{49}
V. B. Svetovoy, P. J.~van Zwol,  G. Palasantzas,  and
J. Th. M. De Hosson,
Optical properties of gold films and the Casimir force,
Phys. Rev. B {\bf 77}, 035439 (2008).
\bibitem{50}
C.\ C.\ Speake and C.\ Trenkel,
Forces between Conducting Surfaces due to Spatial Variations
of Surface Potential,
Phys. Rev. Lett. {\bf 90}, 160403 (2003).
\bibitem{51}
R.~O.~Behunin, D.~A.~R.~Dalvit, R.{\ }S.\ Decca, C.\ Genet,
I.\ W.\ Jung, A.\ Lambrecht, A.\ Liscio, D.~L\'{o}pez,
S.~Reynaud, G.\ Schnoering, G.\ Voisin, and Y.\ Zeng,
Kelvin probe force microscopy of metallic surfaces used
in Casimir force measurements,
Phys. Rev. A {\bf 90}, 062115 (2014).
\bibitem{53a}
C.~D.~Fosco, F.~C.~Lombardo, and F.\ D.\ Mazzitelli,
Proximity force approximation for the Casimir energy as
a derivative expansion,
Phys. Rev. D {\bf 84}, 105031 (2011).
\bibitem{52}
G.~Bimonte, T.~Emig, R.\ L.\ Jaffe, and M.\ Kardar,
Casimir forces beyond the proximity force approximation,
Europhys. Lett. {\bf 97}, 50001 (2012).
\bibitem{53}
G.~Bimonte, T.~Emig, and M.~Kardar,
Material dependence of Casimir force: gradient expansion
beyond proximity,
Appl. Phys. Lett. {\bf 100}, 074110 (2012).
\bibitem{54}
G.~Bimonte,
Going beyond PFA: A precise formula for the sphere-plate Casimir
force,
Europhys. Lett. {\bf 118}, 20002 (2017).
\bibitem{55}
M.~Hartmann, G.-L.~Ingold, and P.~A.~Maia Neto,
Plasma versus Drude Modeling of the Casimir Force: Beyond the
Proximity Force Approximation,
Phys. Rev. Lett. {\bf 119}, 043901 (2017).
\bibitem{56}
A. O. Sushkov, W. J. Kim, D. A. R. Dalvit, and S. K. Lamoreaux,
Observation of the thermal Casimir force,
Nat. Physics {\bf 7}, 230 (2011).
\bibitem{57}
V.{\ }B.\ Bezerra, G.{\ }L.\ Klimchitskaya, U.\ Mohideen,
V.{\ }M.\ Mostepanenko, and C.\ Romero,
Impact of surface imperfections on the Casimir force for
lenses of centimeter-size curvature radii,
Phys. Rev. B {\bf 83}, 075417 (2011).
\bibitem{58}
M. Liu, J. Xu,
G.~L.~Klimchitskaya, V.\ M.\ Mostepanenko, and U.\ Mohideen,
Examining the Casimir puzzle with an upgraded AFM-based technique
and advanced surface cleaning,
Phys. Rev. B {\bf 100}, 081406(R) (2019).
\bibitem{59}
F. J. Giessibl, Advances in atomic force microscopy,
Rev. Mod. Phys. {\bf 75}, 949 (2003).
\bibitem{61}
E.~D.~Palik, O.~J.~Glembocki, I.~Heard, P.\ S.\ Burno, and L.\ Tenerz,
Etching roughness for (100) silicon surfaces in aqueous KOH,
J. Appl. Phys. {\bf 70}, 3291 (1991).
\bibitem{60}
E.-L.~Florin, M. Rief, H. Lehmann, M. Ludwig, C. Dornmair,
V. T. Moy, and H. E. Gaub,
Sensing specific molecular interactions with the atomic force
microscope,
Biosens. Bioelectron. {\bf 10}, 895 (1995).
\bibitem{60a}
Y.-J.~Chen, W.~K.~Tham, D.~E.~Krause, D.~L\'opez,
 E.~Fischbach, and R.~S.~Decca,
Stronger Limits on Hypothetical Yukawa Interactions in the
30--8000~Nm Range,
Phys. Rev. Lett. {\bf 116}, 221102 (2016).
\bibitem{60b}
J.~L.~Garrett, D.~A.~T.~Somers, and J.\ N.\ Munday,
Measurement of the Casimir Force between Two Spheres,
Phys. Rev. Lett. {\bf 120}, 040401 (2018).
\bibitem{62}
R. R. Sowell, R. E. Cuthrell, D. M. Mattox, and R. D. Bland,
Surface cleaning by ultraviolet radiation,
J. Vac. Sci. Technol. {\bf 11}, 474 (1974).
\bibitem{63}
J.~R.~Vig, UV/ozone cleaning of surfaces,
J. Vac. Sci. Technol. A {\bf 3}, 1027 (1985).
\bibitem{64}
N.~S.~McIntyre, R.~D.~Davidson, T.~L.~Walzak, R.\ Williston,
M.\ Westcott, and A.\ Pekarsky,
Uses of ultraviolet/ozone for hydrocarbon removal: Applications
to surfaces of complex composition or geometry,
J. Vac. Sci. Technol. A {\bf 9}, 1355 (1991).
\bibitem{65}
D.~E.~King,
Oxidation of gold by ultraviolet light and ozone at 25\,$^{\circ}$C,
J. Vac. Sci. Technol. A {\bf 13}, 1247 (1995).
\bibitem{66}
A.~Krozer and M.~Rodahl,
X-ray photoemission spectroscopy study of UV/ozone oxidation of Au under
ultrahigh vacuum conditions,
J. Vac. Sci. Technol. A {\bf 15}, 1704 (1997).
\bibitem{67}
S.~R.~Koebley, R.~A.~Outlaw, and R.\ R.\ Dellwo,
Degassing a vacuum system with {\it in-situ} UV radiation,
J. Vac. Sci. Technol. A {\bf 30}, 060601 (2012).
\bibitem{68}
{\it Handbook of Adhesive Technology}, eds. A.\ Pizzi and
K.\ L.\ Mittal, 2nd Edition (Marcel Dekker, New York, 2003).
\bibitem{69}
H.~L\"{u}th,
{\it Surfaces and Interfaces of Solid Materials}
(Springer, Berlin, 2015).
\bibitem{70}
W.~R.~Smythe,
{\it Electrostatics and Electrodynamics}
(McGraw-Hill, New York, 1950).
\bibitem{71}
S.~G.~Rabinovich, {\it Measurement Errors and Uncertainties:
Theory and Practice} (AIP Press, Springer, New York, 2000).
\bibitem {72}
{\it Handbook of Optical Constants of Solids},
ed. E.~D.~Palik (Academic, New York, 1985).
\bibitem{73}
V.~M.~Mostepanenko,
How to confirm and exclude different models of material
 properties in the Casimir effect,
J. Phys.: Condens. Matter {\bf 27}, 214013 (2015).
\bibitem{74}
G.~Bimonte, G.~L.~Klimchitskaya, and V.~M.~Mostepanenko,
Universal experimental test for the role of free charge carriers
 in the thermal Casimir effect within a micrometer separation range,
 Phys. Rev. A {\bf 95}, 052508 (2017).
\bibitem{75}
R.~Sedmik and P.~Brax,
Status Report and first Light from Cannex: Casimir Force
  Measurements between flat parallel Plates,
J. Phys.: Conf. Ser. {\bf 1138}, 012014 (2018).
\bibitem{76}
G.~L. Klimchitskaya, V.~M. Mostepanenko, R.~I.~P.~Sedmik,
and H.\ Abele,
Prospects for Searching Thermal Effects, Non-Newtonian Gravity and
Axion-Like Particles: CANNEX Test of the Quantum Vacuum,
Symmetry {\bf 11}, 407 (2019).
\bibitem{77}
G.~Bimonte,
Apparatus to probe the influence of the Mott-Andersen metal-insulator
transition in doped semiconductors on the Casimir effect,
Phys. Rev. A {\bf 99}, 052506 (2019).
\bibitem{78}
G.~L. Klimchitskaya, V.~M. Mostepanenko, and R.~I.~P.~Sedmik,
Casimir pressure between metallic plates out of thermal equilibrium:
Proposed test for the relaxation properties of free electrons,
Phys. Rev. A {\bf 100}, 022511 (2019).
\bibitem{79}
G.~L. Klimchitskaya and V.~M. Mostepanenko,
Casimir energy and pressure for magnetic metal films,
Phys. Rev. B {\bf 94}, 045404 (2016).
\end{thebibliography}
\end{document}